\documentclass[singlecolumn,epjc3]{svjour3}
\smartqed  
\RequirePackage{graphicx}
\journalname{Eur. Phys. J. C}

\begin{document}

\title{Generalized model for anisotropic compact stars}

\author{S.K. Maurya\thanksref{e1,addr1}
\and Y.K. Gupta\thanksref{e2,addr2}
\and Saibal Ray\thanksref{e3,addr3}
\and Debabrata Deb\thanksref{e4,addr4}.}

\thankstext{e1}{e-mail: sunil@unizwa.edu.om}
\thankstext{e2}{e-mail: kumar$001947$@gmail.com}
\thankstext{e3}{e-mail: saibal@associates.iucaa.in}
\thankstext{e4}{e-mail: d.deb32@gmail.com}

\institute{Department of Mathematical and Physical Sciences,
College of Arts and Science, University of Nizwa, Nizwa, Sultanate
of Oman\label{addr1} \and Department of
Mathematics, Raj Kumar Goel Institute of Technology,
Ghaziabad, Uttar Pradesh, India\label{addr2} \and Department of Physics,
Government College of Engineering and Ceramic Technology, Kolkata
700010, West Bengal, India\label{addr3} \and Department of
Physics, Indian Institute of Engineering Science and Technology,
Shibpur, Howrah 711103, West Bengal, India\label{addr4}}

\date{Received: date / Accepted: date}

\maketitle

\begin{abstract}
In the present investigation an exact generalized model for anisotropic compact stars of embedding
class one is sought for under general relativistic background. The generic solutions
are verified by exploring different physical aspects, viz. energy conditions, mass-radius relation,
stability of the models, in connection to their validity. It is observed that the model present here
for compact stars is compatible with all these physical tests and thus physically acceptable as far as
the compact star candidates $RXJ~1856-37$, $SAX~J~1808.4-3658~(SS1)$ and $SAX~J~1808.4-3658~(SS2)$ are concerned.
\end{abstract}


\maketitle

\section{Introduction}
The studies on anisotropic compact stars remain always a topic of great interest in the relativistic astrophysics. The detailed works of several scientists~\cite{Bowers1974,Ruderman1972,Schunck2003,Herrera1997,Ivanov2002} make our understanding clear about the highly dense spherically symmetric fluid spheres having  pressure anisotropic in nature. Usually anisotropy arises due to presence of mixture of fluids of different types, rotation, existence of superfulid, presence of magnetic field or external field and phase transition etc. According to Ruderman~\cite{Ruderman1972} for the high density $(> {{10}^{15}} gm/{{cm}^{3}})$ anisotropy is the inherent nature of nuclear matters and their interactions are relativistic. In this connection some other works on the anisotropic compact star models can be looked in to the following Refs.~\cite{Mak2003,Usov2004,Varela2010,Rahaman2010,Rahaman2011,Rahaman2012a,Kalam2012,Deb2015}.

Recent study by Randall-Sundram and Anchordoqui-Bergliaffa~\cite{Randall1999,Anchordoqui2000} re-establishes the idea that our 4-dimensional spacetime is embedded in higher dimensional flat space as predicted earlier by Eddington~\cite{Eddington1924}. It is known that the manifold ${V}_{n}$ can be embedded in pseudo-Euclidean space of $m=n(n+1)/2$ dimensions. The class of manifold ${V}_{n}$ which is less than or equal to $m-n=n(n-1)/2$ can be defined as the minimum extra dimension ($p$) of the pseudo-Euclidean
space required for embedding ${V}_{n}$ in ${E}_{m}$. It is to note that when $n=4$, i.e. for the relativistic spacetime ${V}_{4}$, the corresponding value of relativistic embedding class $p$ is 6. The values of the same for the plane and spherical symmetric spacetime are respectively 3 and 2. The class of Kerr is 5~\cite{Kuzeev1980} whereas the class of Schwarzschild's interior and exterior solutions are respectively 2 and 1 and the same for the Friedman-Robertson-Lema{\^i}tre spacetime~\cite{R1933} is 1. In some of our previous works~\cite{M1,M2,M3,M4} we have successfully discussed different stellar models under the embedding class 1.

In this paper utilizing embedding class 1 metric we have attempted to study an anisotropic spherically symmetric stellar model. In this investigation we have assumed that the metric potential $\nu=n\ln  \left(1+{{\it Ar}}^{2} \right) +\ln~B$, where $n \geq 2$. The reasons for the choice of $n \geq 2$ are as follows:\\
(i) with the choice of $n=0$ the space-time becomes flat;\\
(ii) with the choice of $n=1$ this becomes the famous Kohlar-Chao solution~\cite{Kohlar1965};\\
(iii) with the choice of $n=2$, as velocity of sound is not decreasing so we will get a solution which is not well behaved.

We have, therefore, studied our proposed model varying $n=3$ to $n=10000$ and discussed the physical properties of the system for these values of $n$. In the later part one may find from Table~1 that for $n \geq 10$, the product $nA$ becomes almost a constant, say $C$. This situation shows that for large values of $n$ (say infinity) we obtain $\nu=C{r}^{2}+\ln~B$, which is the same metric potential ($\nu=2A{r}^{2}+\ln~B$ with $C=2A$) as considered by Maurya et al.~\cite{M1,Maurya2015a}.

Under the above background the outline of the present work is as follows: we provide the Einstein field equations and their solutions in Sect. 2. In the next Sect. 3 the boundary conditions are discussed to find out constants of integration. The Sec. 4 deals with the applications of the solutions to check several physical properties of the model regarding validity with the stellar structure. Some remarks are passed in concluding Sect. 5.

\section{Basic field equations and solutions}
To describe the interior of a static and spherically symmetry object the line element in the Schwarzschild co-ordinate $(x^{a})=(t,r,\theta,\phi)$ can be written as
\begin{equation}
ds^{2}=e^{\nu(r)}dt^{2}-e^{\lambda(r)}dr^{2}-r^{2}\left(d\theta^{2}+\sin^{2}\theta d\phi^{2} \right),\label{eq1}
\end{equation}
where $\lambda$ and $\nu$ are the functions of the radial coordinate $r$.

Now if the spacetime Eq. (\ref{eq1}) satisfies the Karmarkar condition \cite{karmarkar1948}
\begin{equation}
R_{1414}=\frac{R_{1212}R_{3434}+ R_{1224}R_{1334}}{R_{2323}},\label{eq2}
\end{equation}
with $R_{2323}\neq 0$~\cite{pandey1982}, it represents the spacetime of emending class $1$.

For the condition (\ref{eq2}), the line element Eq. (\ref{eq1}) gives the following differential equation
\begin{equation}
\frac{\lambda'\nu'}{1-e^{\lambda}}=-2(\nu''+\nu'^{2})+\nu'^{2}+\lambda'\nu',\label{eq3}
\end{equation}
with $e^{\lambda}\neq 1$.

Solving Eq. (3) we get
\begin{equation}
e^{\lambda}=1+F\nu'^{2}e^{\nu},\label{eq4}
\end{equation}
where $F \neq 0$ is an arbitrary integrating constant.

We are assuming that within the star the matter is anisotropic and the corresponding energy-momentum tensor can be taken in the form
\begin{equation}
T_{\nu}^{\mu}=(\rho+p_r)u^{\mu}u_{\nu}-p_t g_{\nu}^{\mu}+(p_r-p_t)\eta^{\mu}\eta_{\nu},\label{eq5}
\end{equation}
with $ u^{i}u_{j} =-\eta^{i}\eta_j = 1 $ and $u^{i}\eta_j= 0$, the vector $u_i$ being the fluid 4-velocity and $\eta^{i}$ is the space-like vector which is orthogonal to $ u^{i}$. Here $\rho$ is the matter density, $p_r$ is the the radial and $p_t$ is transverse pressure of the fluid in the orthogonal direction to $p_r$.

Assuming $\kappa=8\pi$ with $G=c=1$ (in relativistic geometrized unit) the Einstein field equations are given by
\begin{equation}
\frac{1-e^{-\lambda}}{r^{2}}+\frac{e^{-\lambda}\lambda'}{r}=\kappa\rho,\label{eq6}	
\end{equation}

\begin{equation}
\frac{e^{-\lambda}-1}{r^{2}}+\frac{e^{-\lambda}\nu'}{r}=\kappa\, p_{r},\label{eq7}
\end{equation}

\begin{equation}	
e^{-\lambda}\left(\frac{\nu''}{2}+\frac{\nu'^{2}}{4}-\frac{\nu'\lambda'}{4}+\frac{\nu'-\lambda'}{2r} \right)=\kappa\, p_t.\label{eq8}
\end{equation}

Here we have four equations with five unknowns, namely $\lambda,~\nu,~\rho,~p_r$ and $p_t$ which are to be find out under the proposed model. This immediately prompt us to explore for some suitable relationship between the unknowns or an existing physically acceptable metric potential can be opted for which will help us to overcome the mathematical situation of redundancy.

Therefore, to solve the above set of Einstein field equations let us take the metric co-efficient, $e^{\nu}$, as proposed by Lake \cite{Lake2003}
\begin{equation}
e^{\nu}=B(1+Ar^{2})^{n},\label{eq9}
\end{equation}
where $A$ and $B$ are constants and $n \ge 2$.

\begin{figure}[!htp]\centering
    \includegraphics[width=5.5cm]{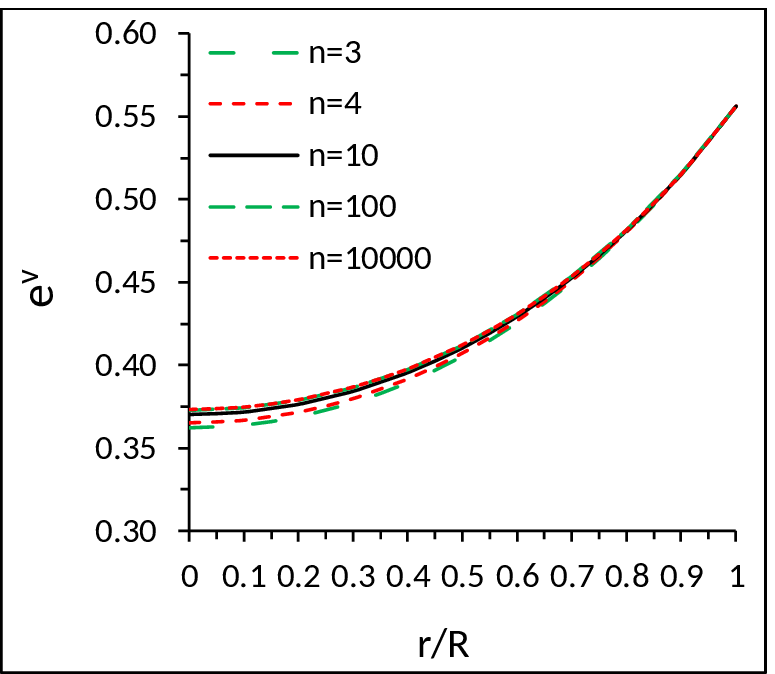}
    \includegraphics[width=5.5cm]{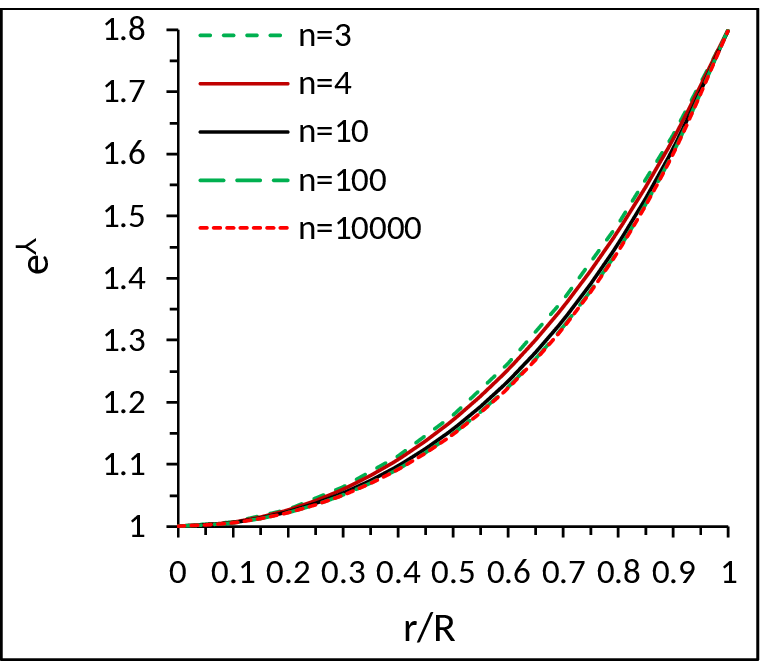}
\caption{Variation of metric functions $e^{\nu}$ and $e^{\lambda}$ with the fractional coordinate $r/R$ for $RXJ~1856-37$}
    \label{Fig1}
\end{figure}

Solving Eqs. (4) and (9) we obtain
\begin{equation}
e^{\lambda}=[1+D\,Ar^2 (1+Ar^2)^{(n-2)}],\label{eq10}
\end{equation}
where $D=4n^2\,A\,B\,F.$

Now using Eqs. (\ref{eq6})-(\ref{eq8}), (\ref{eq9}) and (\ref{eq10}) we obtain the expression for $\rho$, ${{p}_{r}}$ and ${{p}_{t}}$ as
\begin{equation}
\frac{\kappa\rho}{A}=\frac{D\,(1+Ar^2)^n\,[3+(2n-1)A^2r^4+2Ar^2+2\,n\,Ar^2+D\,Ar^2(1+Ar^2)^n]}{[1+A^2r^4+2\,Ar^2\,+D\,Ar^2\,(1+Ar^2)^n]^2},\label{eq11}
\end{equation}

\begin{figure}[!htp]\centering
    \includegraphics[width=5.5cm]{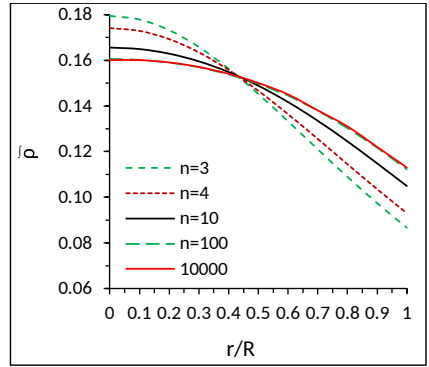}
\caption{Variation of effective density, $\tilde \rho=\rho/(n A)$, with the fractional coordinate $r/R$ for $RXJ~1856-37$}
    \label{Fig3}
\end{figure}

\begin{equation}
\frac{\kappa\, p_r}{A}=\frac{2\,n\,(1+Ar^2)-D\,(1+Ar^2)^n}{[1+A^2r^4+2\,Ar^2\,+D\,Ar^2\,(1+Ar^2)^n]},\label{eq12}
\end{equation}

\begin{equation}
\frac{\kappa p_t}{A}=\frac{(1+Ar^2)\,[2\,n\,(1+Ar^2)+n^2\,Ar^2\,(1+Ar^2)+D\,(Ar^2-1)\,(1+Ar^2)^n]}{[1+A^2r^4+2\,Ar^2\,+D\,Ar^2\,(1+Ar^2)^n]^2},\label{eq13}
\end{equation}

\begin{figure}[!htp]\centering
    \includegraphics[width=5cm]{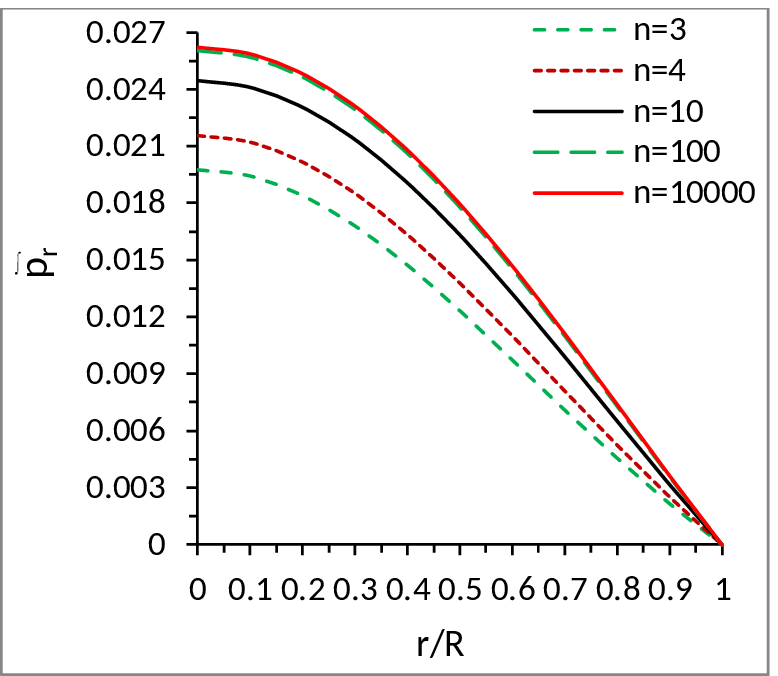}
    \includegraphics[width=5cm]{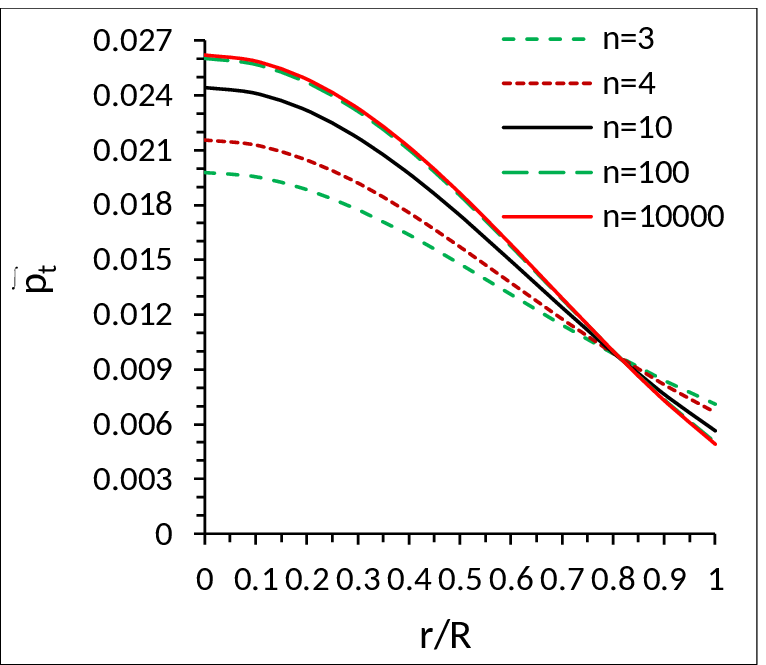}
\caption{Variation of effective radial pressure, $\tilde p_r=p_r/(nA)$, (left panel) and effective tangential pressure, $\tilde p_t=p_t/(nA)$, (right panel) with the fractional coordinate $r/R$ for $RXJ~1856-37$}
    \label{Fig4}
\end{figure}

and the anisotropic factor $\Delta$ is obtained as
\begin{equation}
 \Delta=\frac{A^2r^2\,[n^2\,f^2-2n\,f\,(f+D\,f^n)+D\,f^n(2+2\,Ar^2+D\,f^n)]}{\kappa\,[1+A^2r^4+2\,Ar^2\,+D\,Ar^2\,(1+Ar^2)^n]^2},\label{eq14}
\end{equation}
where $f=(1+Ar^2)$. The profiles of the metric functions, effective density, the effective radial and tangential pressures, the anisotropic factor are respectively shown in Figs. 1-4.

From Fig. 4 we also find that for our system the anisotropic factor is minimum at the centre and it is maximum at the surface as proposed by Deb et al.~\cite{Deb2016} for the anisotropic stellar model. However the anisotropy factor is zero for all radial distance $r$ if and only if $A=0$. This implies that in the absence  of anisotropy the radial and transverse pressures and density become zero. Also in this case the metric turns out to be flat.

\begin{figure}[!htp]\centering
    \includegraphics[width=5.5cm]{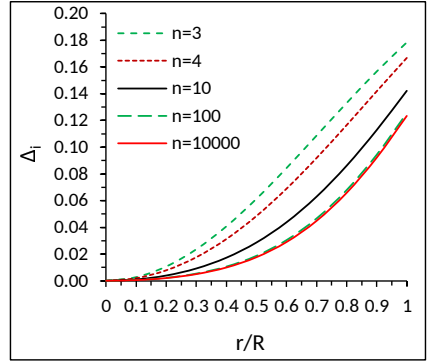}
\caption{Variation of anisotropic factor ($\Delta_i=\kappa\Delta/A$) with the fractional coordinate $r/R$ for $RXJ~1856-37$}
    \label{Fig5}
\end{figure}

\section{Boundary conditions to determine the constants}
For fixing the values of the constants, we match our interior space-time to the exterior Schwarzschild line element given by
\begin{equation}
ds^{2}=\left(1-\frac{2m}{r}\right)dt^{2}-\left(1-\frac{2m}{r}\right)^{-1}dr^{2}-r^{2}(d\theta^{2}+\sin^{2}\theta d\phi^{2}).\label{eq15}
\end{equation}

Outside the event horizon $r>2m$, $m$ being the mass of the black hole. Also the radial pressure $p_r$ must be finite and positive at the centre of the star which must vanish at the surface $r = R$~\cite{Misner1964}. Then $p_{r}(R)$ = 0 gives:
\begin{equation}
D=4n^2\,A\,B\,F=2\,n\,(1+A\,R^2)^{1-n}.\label{eq16}
\end{equation}

This yields the radius of star as
\begin{equation}
R=\sqrt{\frac{(2\,n\,A\,B\,F)^{1/(1-n)}-1}{A}}, \label{eq17}
\end{equation}
using the continuity of the metric coefficient $e^{\nu},e^{\lambda}$ and $\frac{\partial g_{tt}}{\partial r}$ (matching of second fundamental form at $r=R$ is same as radial pressure should be zero at $r=R$) across the boundary we get the following three equations
\begin{equation}
1-\frac{2M}{R}=B(1+A R^{2})^{n},\label{eq18}
\end{equation}

\begin{equation}
\left(1-\frac{2M}{R}\right)^{-1}=1+D\,AR^2 (1+AR^2)^{(n-2)},\label{eq19}
\end{equation}

\begin{equation}
\frac{2M}{R^{3}}=2\,n\,B\,A(1+AR^{2})^{n-1}.\label{eq20}
\end{equation}

Solving Eq. (16) and Eqs. (18)-(20), in terms of mass of radius of the compact star, we obtained the expressions for $B$, $F$ and $M$ as
\begin{equation}
B=\frac{\left(1+AR^2\right)^{1-n}}{1+(1+2\,n)AR^2},\label{eq21}
\end{equation}

\begin{equation}
F=\frac{1+(1+2\,n)AR^2}{2\,n\,A},\label{eq22}
\end{equation}

\begin{equation}
\frac{M}{R}=\frac{n\,AR^2}{1+AR^2(1+2n)}.\label{eq23}
\end{equation}

However, the value of arbitrary constant $A$ is determined by using the density of star at the surface, i.e. $\rho_{r=R} = \rho_{R}$ as
\begin{equation}
A=\frac{\kappa\rho_{R}[1+A^2R^4+2\,AR^2\,+D\,AR^2\,(1+AR^2)^n]^2}{D\,(1+AR^2)^n\,[3+(2n-1)A^2R^4+2AR^2+2\,n\,AR^2+D\,AR^2(1+AR^2)^n]}.\label{eq24}
\end{equation}

The gradients of density and radial pressure (by taking $x=Ar^2$) as
\begin{equation}
\frac{\kappa}{A}\,\frac{d\rho}{dr}=-\frac{D\,f^n\,[-2\,x^3+10+5D\,f^n+6\,x^2+3D\,x^2\,f^n+\rho_1(x)-2\,n^2\,x\,\rho_2(x)]}{[1+x^2+2\,x\,+D\,x\,(1+x)^n]^3},\label{eq25}
\end{equation}

\begin{equation}
\frac{\kappa}{A}\,\frac{dp_r}{dr}=-\frac{2\,D\,n^2\,x\,f^n-D\,f^n\,[2f+D\,f^n]+n\,[2+2x^2+3D\,f^n+4\,x+D\,x\,f^n)]}{[1+x^2+2\,x\,+D\,x\,(1+x)^n]^2},\label{eq26}
\end{equation}
where\\
$\rho_1(x)=x\,[18+4\,D\,(1+x)^n+D^2\,(1+x)^(2\,n)]+n\,(x-1)[5+5\,x^2+x\,(10-3\,D\,(1+x)^n)]$,\\ $\rho_2(x)=[1+x^2+2x-D\,x\,(1+x)^n]$, $f=(1+x)$.

\section{Physical features of the anisotropic models}
In this section different physical features of the compact stars will be discussed.

\subsection{Energy Conditions}
To satisfy energy conditions i.e null energy condition (NEC), weak energy
condition (WEC) and strong energy condition (SEC) the anisotropic fluid spheres must be consistent with the following inequalities simultaneously inside the stars given as
\begin{equation}
NEC: \rho\geq 0,\label{eq27}
\end{equation}

\begin{eqnarray}
WEC: \rho- {{p}_{r}} \geq  0 \hspace{0.1cm}({WEC}_{r}) \hspace{0.2cm} and
     \hspace{0.2cm} \rho - p_t \geq  0 \hspace{0.1cm}({WEC}_{t}),\label{eq28}
\end{eqnarray}

\begin{equation}
SEC: \rho-p_r-2{{p}_{t}} \geq  0.\label{eq29}
\end{equation}

From Fig. 5 it is clear that the energy conditions are satisfied in the interior of the compact stars simultaneously.

\begin{figure}[!h]\centering
	\includegraphics[width=4.6cm]{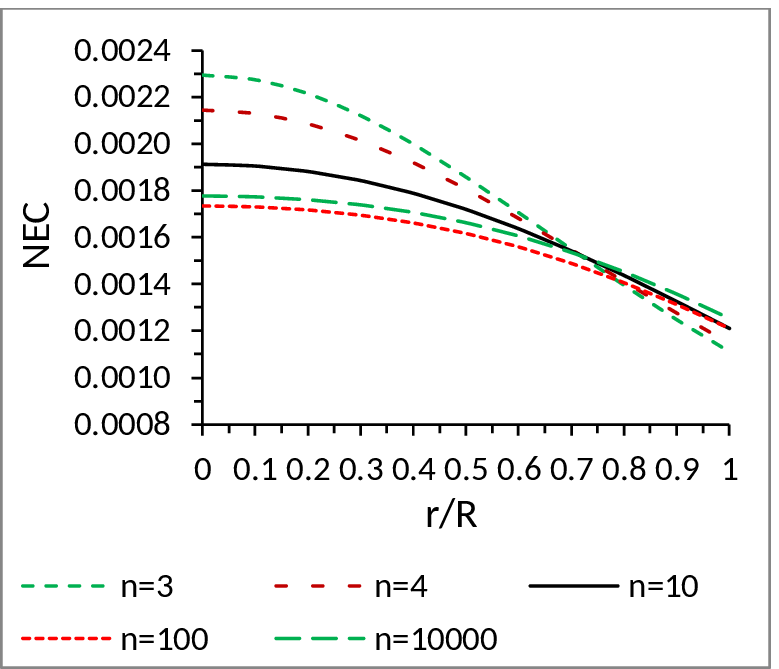}
\includegraphics[width=4.6cm]{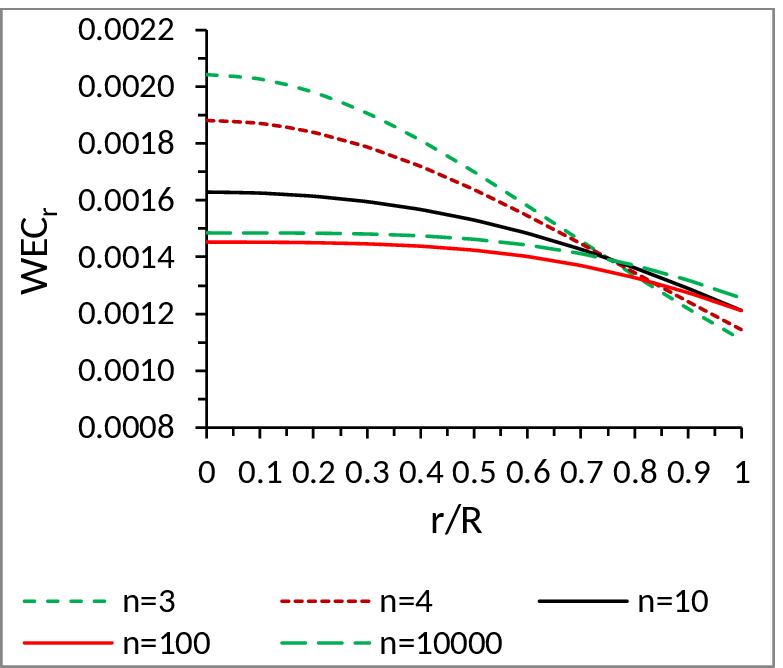}
     \includegraphics[width=4.6cm]{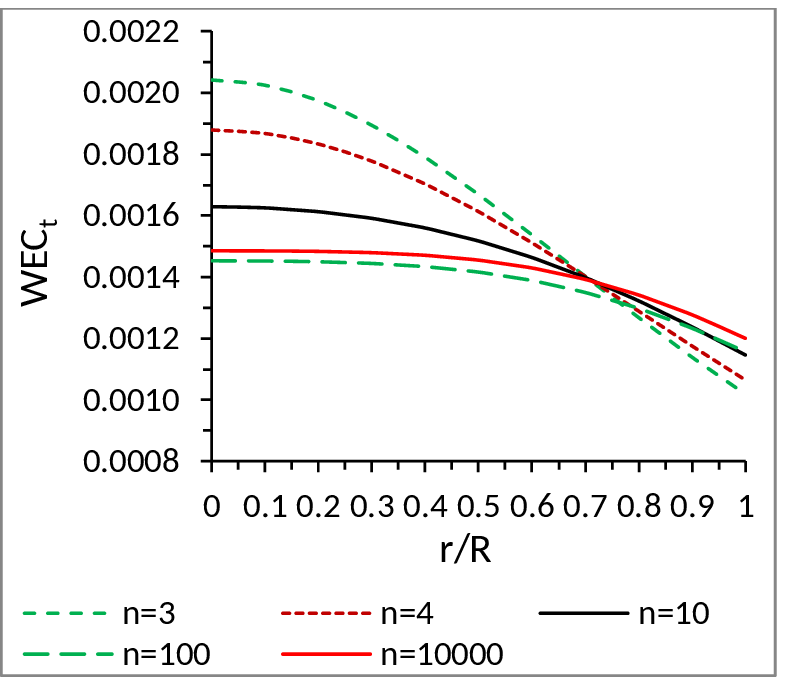}
     \includegraphics[width=4.6cm]{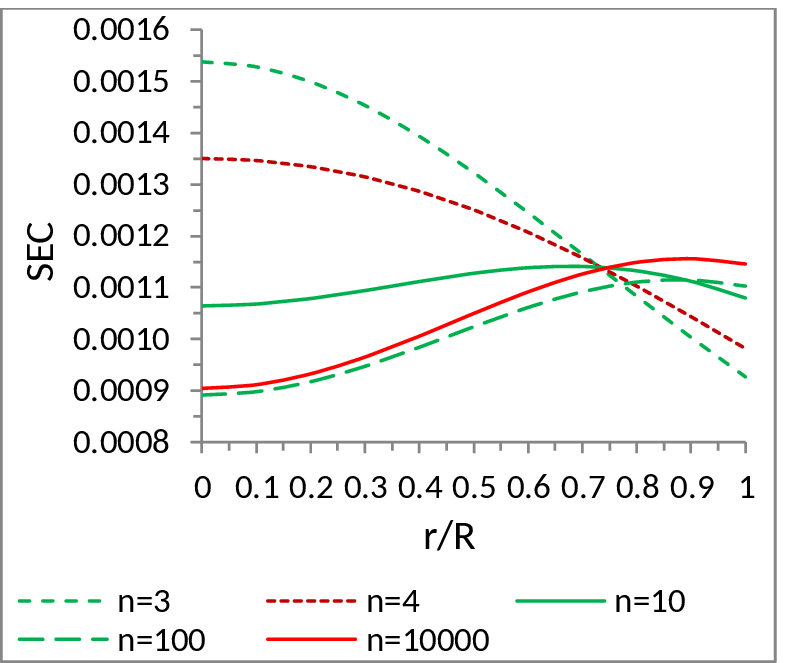}
	\caption{Variation of energy condition with respect to fractional radius ($r/R$) for $RXJ~1856-37$: (i) NEC (top left), (ii) $WEC_r$ for (top right), (iii) $WEC_t$ (bottom left), (iv) SEC (bottom right) }
\label{Fig6}
\end{figure}

\subsection{Mass-radius relation}
For the physical validity of the model according to Buchdahl~\cite{Buchdahl1959} the mass to radius ratio for perfect fluid should be $2M/R \leq 8/9$ which was later proposed in a more generalized expression by Mak and Harko~\cite{Mak2003}.

The Effective mass of the compact star is obtained as
\begin{equation}
M_{eff}=\frac{\kappa}{2}\int_0^{R}\rho\, r^{2}dr=\frac{n\,AR^3}{1+AR^2(1+2n)}. \label{eq30}
\end{equation}

In Fig.~6 variation of maximum mass with respect to corresponding radius of the compact stars with the different values of $n$ have shown. We have also plotted in Fig.~7 variation of maximum values of $2M/R$ for the different values of $n$. We found throughout the study for the different values of n our system is valid with the Buchdahl conditions~\cite{Buchdahl1959} which is also clear from Fig.~7.

\begin{figure}[!htp]\centering
    \includegraphics[width=8.5cm]{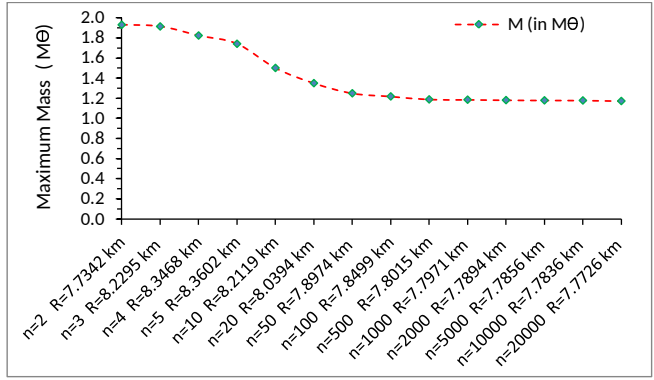}
\caption{Variation of maximum mass $M$ (in km) with respect to corresponding radius $R$ (in km) for different values of $n$ }
    \label{Fig8}
\end{figure}

\begin{figure}[!htp]\centering
    \includegraphics[width=8.5cm]{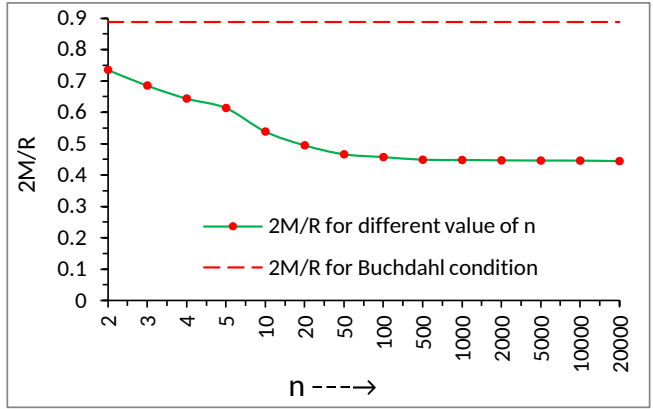}
\caption{Variation of maximum values of $\frac{2M}{R}$ for the different values of $n$ of the anisotropic compact stars}
    \label{Fig9}
\end{figure}

The compactification factor of the stars is obtained as
\begin{equation}
u(r)=\frac{M(r)}{r}=\frac{D\,AR^{2}\,(1+AR^{2})^{n-2}}{2[1+D\,AR^{2}\,(1+AR^{2})^{n-2}]}.\label{eq31}
\end{equation}

The surface redshift $z_s$ with respect to the above compactness (u) is given as
\begin{equation}
z_s=\left(1-2u\right)^{-\frac{1}{2}}-1=\sqrt{1+D\,AR^{2}\,(1+AR^{2})^{n-2}}-1], \label{eq32}
\end{equation}
whose behaviour is shown in Fig.~8 with the fractional coordinate $r/R$ for $RXJ~1856-37$.

\begin{figure}[!htp]\centering
    \includegraphics[width=5.5cm]{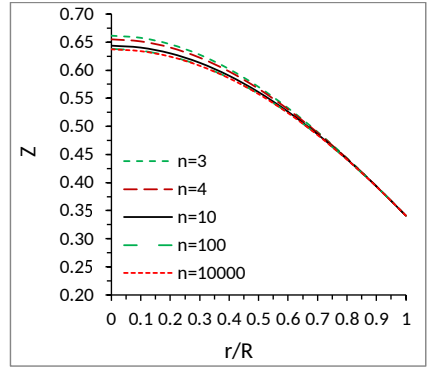}
\caption{Variation of redshift ($Z$) with the fractional coordinate $r/R$ for $RXJ~1856-37$}
    \label{Fig10}
\end{figure}

\subsection{Stability of the model}
In the following sub sections we will try to study the stability of the proposed mathematical model.

\subsubsection{Generalized TOV equation}
Following Tolman~\cite{Tolman1939}, Oppenheimer and Volkoff~\cite{Oppenheimer1939} we want to examine whether our present model is stable under the three forces, viz. gravitational force ($F_g$), hydrostatics force ($F_h$) and anisotropic force ($F_a$) so that the sum of the forces becomes zero for the system to be in equilibrium, i.e.
\begin{equation}
F_g+F_h+F_a=0.\label{eq36}
\end{equation}

The generalized Tolman–Oppenheimer–Volkoff (TOV) equation~\cite{Leon1993,Varela2010} for our system takes form as following
\begin{equation}
-\frac{M_G(r)(\rho+p_r)}{r}e^{\frac{\nu-\lambda}{2}}-\frac{dp_r}{dr}+\frac{2}{r}(p_t-p_r)=0,\label{eq33}
\end{equation}
where $M_G(r)$ represents the gravitational mass within the radius $r$, which can derived from the Tolman-Whittaker formula~\cite{Devitt1989} and the Einstein field equations and is defined by
\begin{equation}
M_G(r)=\frac{1}{2}re^{\frac{\lambda-\nu}{2}\nu'}.\label{eq34}
\end{equation}

Plugging the value of $M_G(r)$ in equation $(25)$, we get
\begin{equation}
-\frac{\nu'}{2}(\rho+p_r)-\frac{dp_r}{dr}+\frac{2}{r}(p_t-p_r)=0.\label{eq35}
\end{equation}

Now using Eqs. (\ref{eq16}-\ref{eq18}), the expressions for $F_g$, $F_h$ and $F_a$ can be written as
\begin{equation}
F_g=-\frac{\nu'}{2}(\rho+p_r)=-\frac{2n\,A^2\,r\,[D\,(1-x)\,f^n+n+n\,x^2+2\,n\,x+2\,D\,n\,x\,f^n]}{\kappa\,[1+x^2+2\,x\,+D\,x\,(1+x)^n]^2},\label{eq37}
\end{equation}

\begin{equation}
F_h=\frac{2A^2\,r}{\kappa}\,\frac{2\,D\,n^2\,x\,(1+x)^n-D\,f^n\,[2+2x+D\,f^n]+n\,F_{h1}}{[1+x^2+2\,x\,+D\,x\,(1+x)^n]^2},\label{eq38}
\end{equation}

\begin{equation}
F_a=\frac{2A^2\,r\,[n^2\,f^2-2n\,f\,(f+D\,f^n)+D\,f^n(2\,f+D\,f^n)]}{\kappa\,[1+x^2+2\,x\,+D\,x\,(1+x)^n]^2},\label{eq39}
\end{equation}
where,
$F_{h1}=[2+2x^2+3D\,(1+x)^n+4\,x+D\,x\,(1+x)^n)]$ and $f=(1+x)$. We have shown the behaviour of TOV equation with $n=3-10000$ in Fig.~9 for $RXJ~1856-37$. As far as equilibrium is concerned the plots are satisfactory in their nature.

\begin{figure}[!h]\centering
	\includegraphics[width=4.3cm]{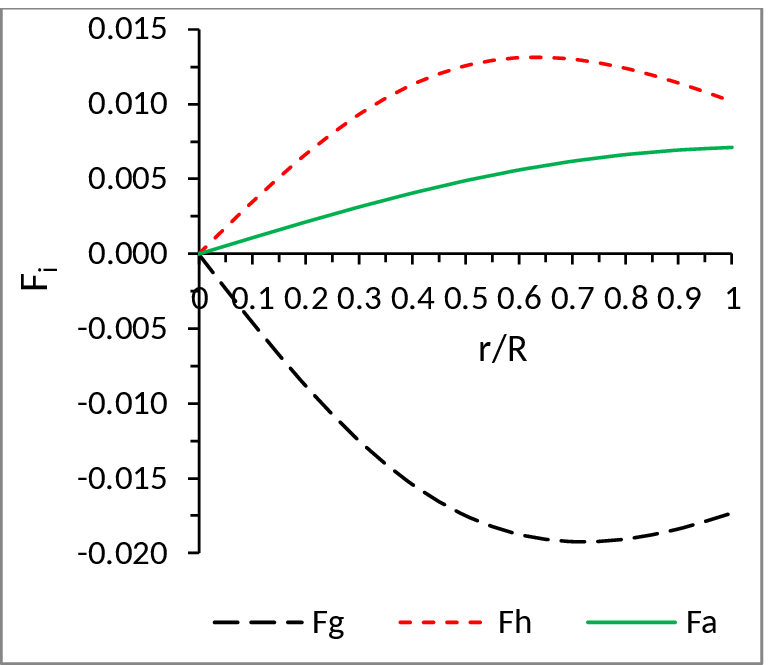}
\includegraphics[width=4.3cm]{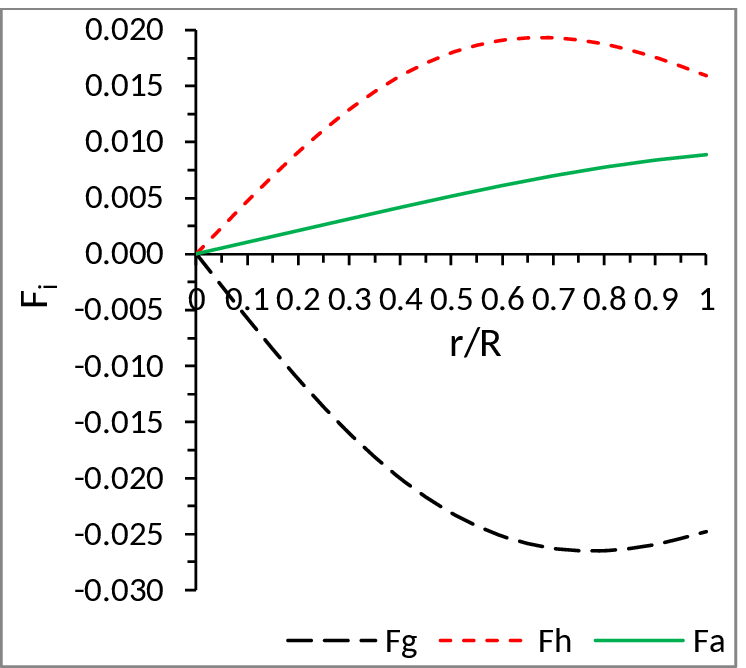}
\includegraphics[width=4.3cm]{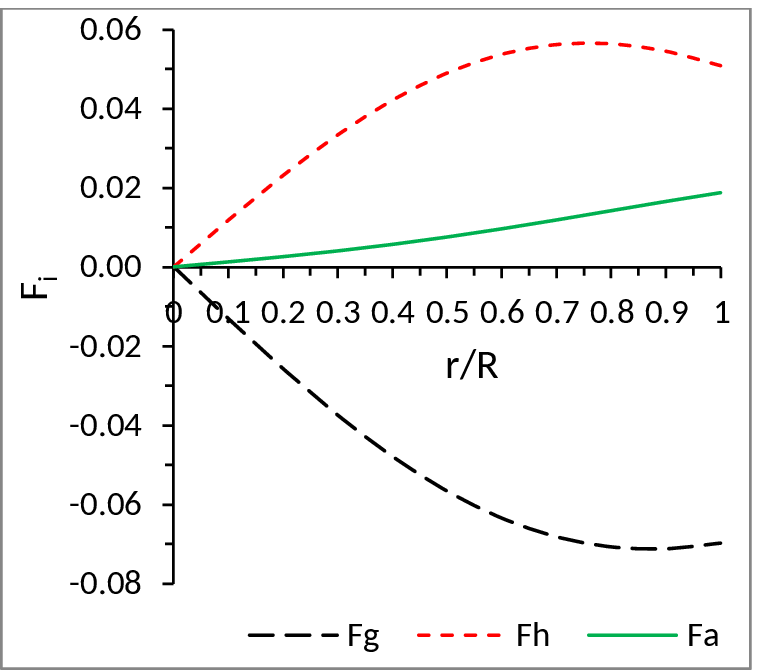}
\includegraphics[width=4.5cm]{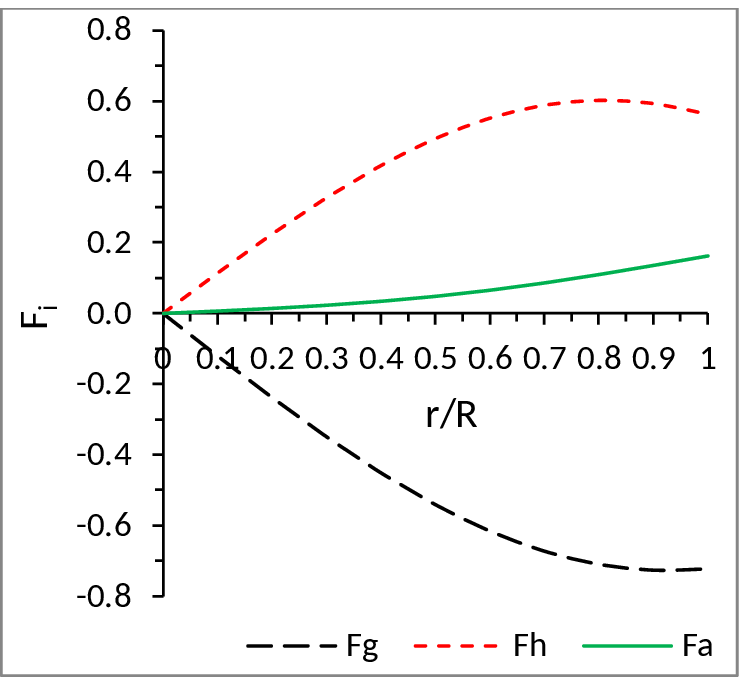}
\includegraphics[width=4.5cm]{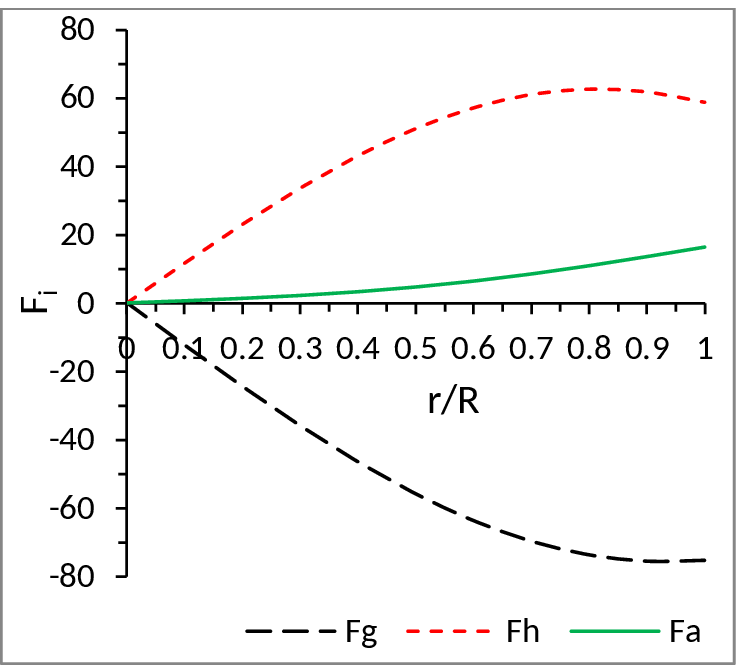}
	\caption{Variation of the different forces with respect to fractional radius ($r/R$) for $RXJ~1856-37$: (i) $n=3$ (top left), (ii) $n=4$ (top middle), (iii) $n=10$ (top right), (iv) $n=100$ (bottom left), (v) $n=10000$ (bottom right)}.
	\label{Fig9}
\end{figure}

\subsubsection{Herrera's cracking concept}
With the help of Herrera's~\cite{Herrera1992} `cracking concept' we try to examine the stability of the proposed configuration. For physical validity the fluid distribution must admit the condition of causality which suggests that square of radial $({{v}_{r}}^2)$ and tangential $({{v}_{t}}^2)$ sound speeds individually must lie within the limit 0 and 1. Also following Herrera~\cite{Herrera1992} and Abreu et al.~\cite{Abreu2007} it can be concluded that for the stable region provided condition is $|{v_{st}}^2- {v_{sr}}^2| \leq 1$ which indicates that for the stable region `no cracking' is the another essential condition. For our system sound velocities are
\begin{equation}
	v_{r}^{2}=\frac{dp_{r}}{d\rho}=\frac{h(x)}{D\,f^n}\left[\frac{2\,D\,n^2\,x\,f^n-D\,f^n\,[2+2x+D\,f^n]+n\,v_{r1}}
{g(x)+3D\,x^2\,f^n+x\,\rho_1(x)+n\,(x-1)\,\rho_2(x)-2\,n^2\,x\,\rho_3(x)]}\right],\label{eq40}
\end{equation}

\begin{equation}
v_{t}^{2}=\frac{dp_{t}}{d\rho}=\frac{h(x)}{D\,f^n}\,\left[\frac{-2\,f\,v_{t1}+\,v_{t2}[1+x^2+2\,x+ D\,x\,f^n]+f\,v_{t3}}{g(x)+3D\,x^2\,f^n+x\,\rho_1(x)+n\,(x-1)\,\rho_2(x)-2\,n^2\,x\,\rho_3(x)]}\,\right],\label{eq41}
\end{equation}

\begin{figure}[!htp]\centering
    \includegraphics[width=5cm]{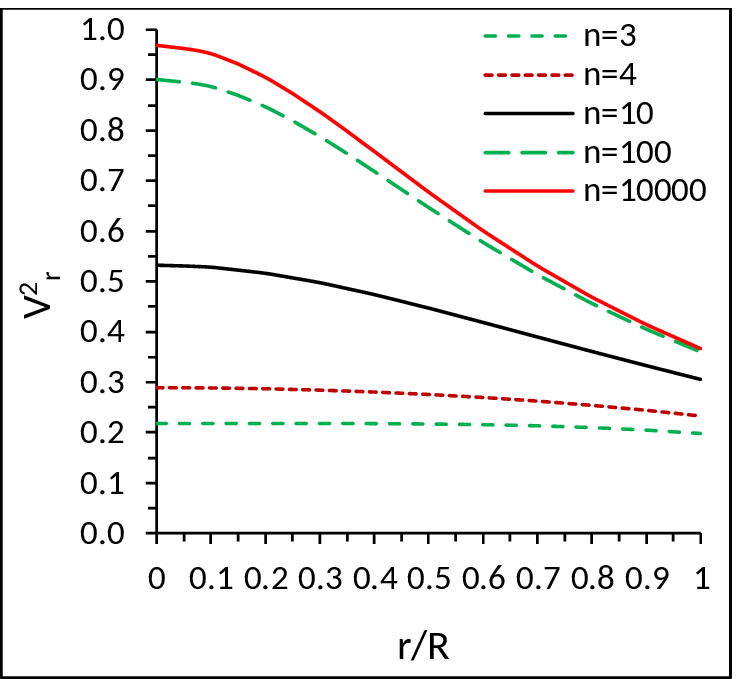} 
    \includegraphics[width=5cm]{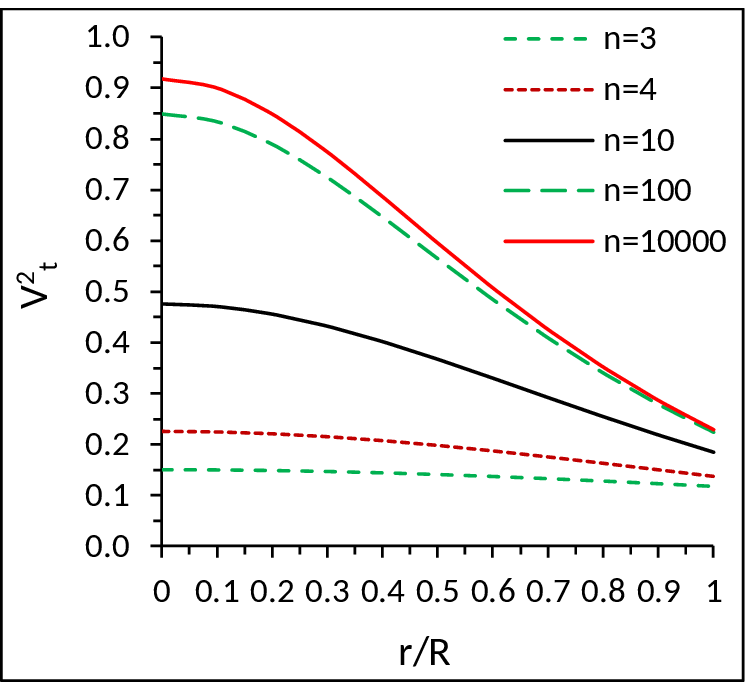}
\caption{Variation of square of radial velocity $V^{2}_r$ and transverse velocity $V^{2}_t$  with the radial coordinate ($r/R$) for $RXJ~1856-37$}
    \label{Fig12}
\end{figure}

\begin{figure}[!htp]\centering
    \includegraphics[width=5cm]{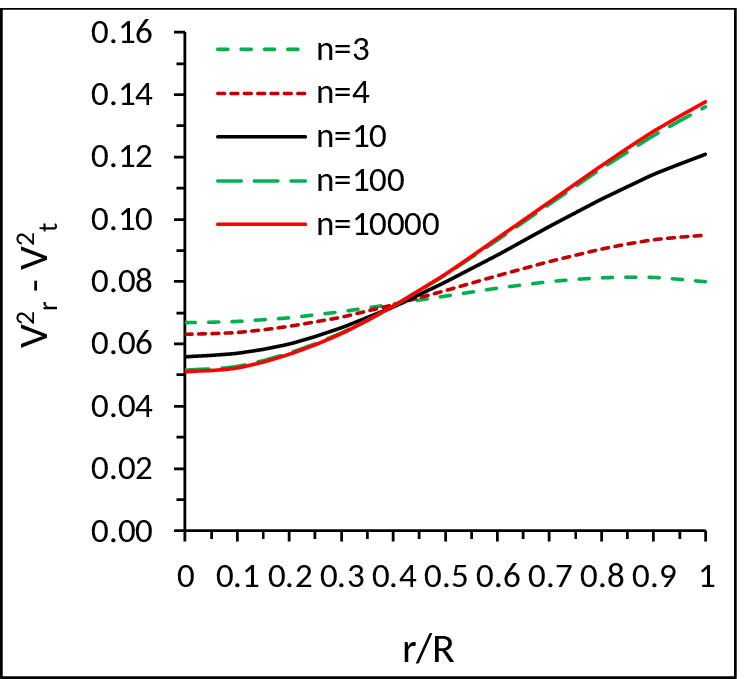} 
    \includegraphics[width=5cm]{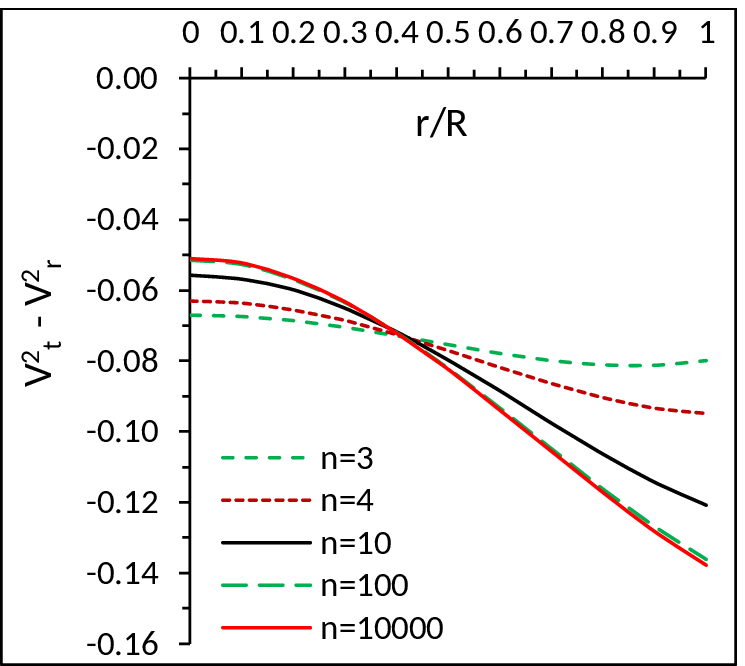}

\caption{Variation of difference of the square of sound velocities with the radial coordinate $r/R$ for $RXJ~1856-37$}
    \label{Fig13}
\end{figure}

where\\
$h(x)=[1+x^2+2\,x+D\,x\,f^n]$; $g(x)=[-2\,x^3+10+5D\,f^n+6\,x^2]$\\
$v_{r1}=[2+2x^2+3D\,f^n+4\,x+D\,x\,f^n)]$;\\
$v_{t1}=2\,n\,f+n^2\,x\,f+D\,(x-1)\,f^n\,[2+D\,f^n+2\,x+D\,n\,x\,f^n]$;\\
$v_{t2}=[2\,n\,f+n^2\, x\,f+D\,(-1 + x)\,f^n]$;\\
$v_{t3}=D\,f^n+n^2 (1+2\,x)+n\,[2+D(-1 + x)\,f^{n-1}]$.

From Figs. 10-11 it is clear that our stellar system satisfies all the conditions said above and hence provides a stable configuration.

\subsubsection{Adiabetic index}
According to Heintzmann and Hillebrandt~\cite{Heintzmann1975} the required condition for the stability of the isotropic compact stars is the adiabatic index, $\gamma > \frac{4}{3}$ in the all interior points of the stars. For our model we have
\begin{equation}
\gamma_r = \frac{\rho+p_r}{p_r}\frac{dp_r}{d\rho}= \frac{2(1+x)\,[D\,(1-x)\,(1+x)^n+n+n\,x^2+\Gamma_{r1}]}{2\,n\,(1+x)-D\,(1+x)^n}\frac{dp_r}{d\rho},
\end{equation}

\begin{equation}
\gamma_t = \frac{\rho+p_t}{p_t}\frac{dp_t}{d\rho}=\frac{n^2\,x\,(1+x)^2+2\,n\,(1+x)[1+x+Dx\,(1+x)^n]+\Gamma_{t1}}{(1+x)\,[(2\,n\,+n^2\,x)\,(1+x)+D\,(-1+x)\,(1+x)^n]}
\frac{dp_t}{d\rho},
\end{equation}
where\\
$\Gamma_{r1}=[2\,n\,x+2\,D\,n\,x\,(1+x)^n]$, $\Gamma_{t1}=D\,(1+x)^n\,[2+2\,x+D\,x\,(1+x)^n]$.

\begin{figure}[!htp]\centering
    \includegraphics[width=5cm]{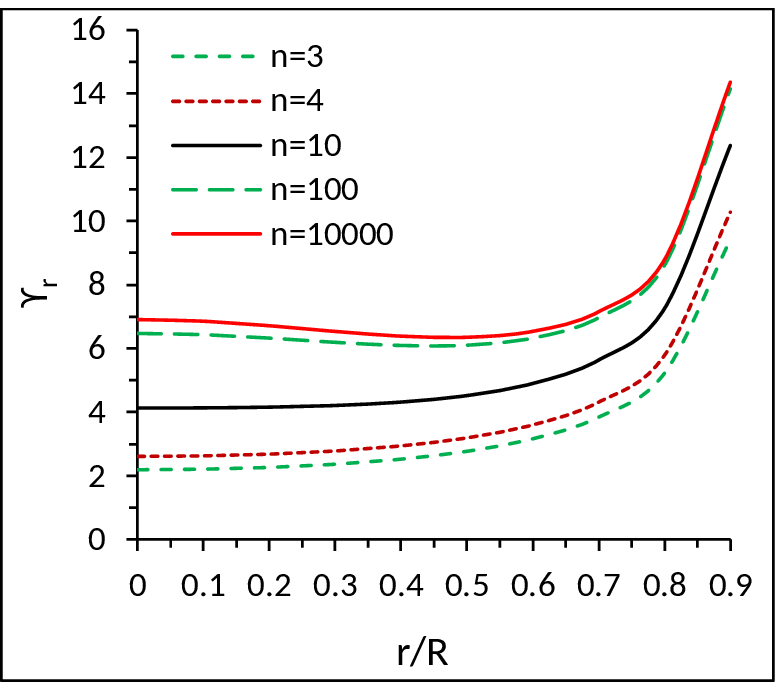} 
    \includegraphics[width=5cm]{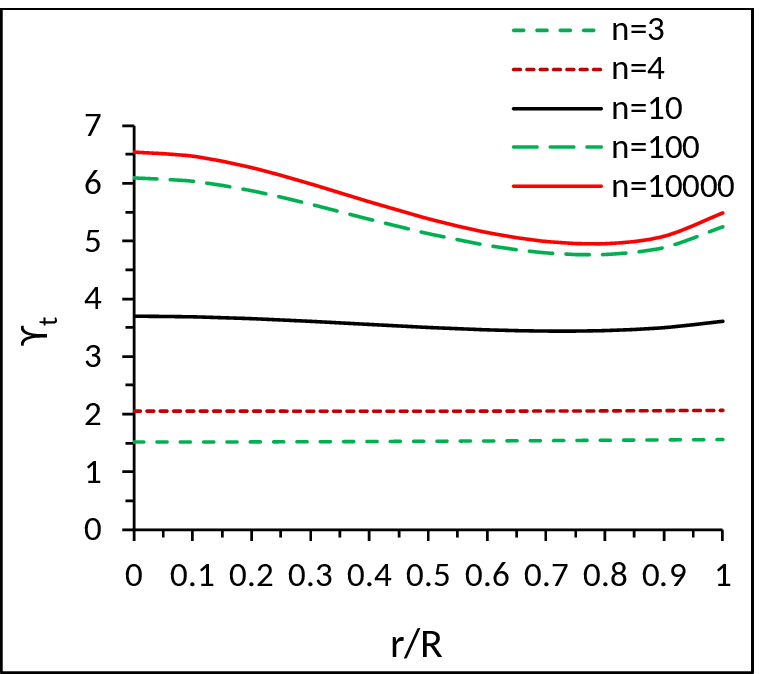}
\caption{Variation of adiabatic index $\gamma_r$ and $\gamma_t$ with the radial coordinate ($r/R$) for $RXJ~1856-37$}
    \label{Fig14}
\end{figure}

From Fig. 12 it is obvious that for all $n$ the values of $\gamma_r,\gamma_t$ are greater than $4/3$ and hence our system is stable.

\section{Discussions and conclusions}
In the present paper we have performed certain investigations on the nature of compact stars by utilizing embedding class one metric. Here an anisotropic spherically symmetric stellar model has been considered. To carried out the investigations we have considered the following assumptions that the metric $\nu=n\ln  \left(1+{{\it Ar}}^{2} \right) +\ln~B$, where $n \geq 2$. The reasons for such consideration on $n \geq 2$ are already have mentioned in the introductory part and are as follows: (i) with the choice of $n=0$ the spacetime becomes flat as Minkowski type; (ii) with the choice of $n=1$ this becomes the famous Kohlar-Chao solution~\cite{Kohlar1965} and (iii) with the choice of $n=2$, the spacetime does not provide well behaved solutions.

Under the above circumstances, therefore, we have studied our proposed model for variation of $n=3$ to $n=10000$. We find from Table~1 that for $n \geq 10$, the product $nA$ becomes almost a constant (say $C$). Thus we can conclude that for the large values of $n$ (say infinity) we have the metric potential $\nu=C{r}^{2}+\ln~B$ as considered by Maurya et al.~\cite{M1,Maurya2015a} in their previous literature. This result therefore helps us in turn to explore behaviour of the mass and radius of the spherical stellar system as can be observed from Fig. 6. We have shown here variation of maximum mass $M$ (in km) with respect to corresponding radius $R$ (in km) for different value of $n$. The profile is very indicative which shows that up to $n=3$ and $R=8.2295$ km maximum mass feature is roughly steady and after $n > 3$ it gradually decreases. However, after $n=500$ and $R=7.8015$ km the maximum mass again acquires almost a steady feature. In Fig. 7 the Buchdahl condition, i.e. mass-radius relation regarding stable configuration of the stellar system has been shown to be satisfactorily followed.

The main features of the present work therefore can be highlighted for the nature of compact stars as follows:

(1) The stars are anisotropic in their configurations unless $p_r \neq p_t$. The radial pressure $p_r$ vanishes but tangential pressure
$p_t$ does not vanish  at the boundary $r = R$ (radius of the star). However, the radial pressure is equal to the tangential
pressure at the centre of the fluid sphere. The anisotropy factor is zero for all radial distance $r$ if and only if $A=0$. This implies that in the absence  of anisotropy the radial and transverse pressures and density become zero. Also the metric turns out to be flat. 

(2) To solve the Einstein field equations we consider the metric co-efficient, $e^{\nu}=B(1+Ar^{2})^{n}$ as proposed by Lake~\cite{Lake2003} and hence the spacetime of the interior of the compact stars can be described by Lake metric.

(3) We observe from Fig.~\ref{Fig5} that for our system the anisotropic factor is minimum at the centre and maximum at the surface as proposed by Deb et al.~\cite{Deb2016}. However, the anisotropy factor is zero for all radial distance $r$ if and only if $A=0$. This implies that in the absence of anisotropy the pressures and density become zero which in turn makes the metric to be flat.

(4) The energy conditions are fulfilled as can be seen from Fig. 5 under variation of $n$.

(5) We have discussed about the stability of the model by applying (i) the TOV equation, (ii) the Herrera cracking concept, and (iii) the adiabatic index of the interior of the star. It can be observed that stability of the model has been attained surprisingly in our model (see Figs. 9-11).

(6) The surface redshift analysis for our case shows that for the compact star $RXJ~1856-37$ this turns out to be $0.65$ as maximum value. In the isotropic case and in the absence of the cosmological constant it has been shown that $Z \leq 2$ \cite{Buchdahl1959,Straumann1984,Boehmer2006} whereas B{\"o}hmer and
Harko~\cite{Boehmer2006} argued that for an anisotropic star in the presence of a cosmological constant the surface redshift must obey the general restriction $Z \leq 5$, which is consistent with the bound $Z \leq 5.211$ as obtained by Ivanov~\cite{Ivanov2002}. Therefore, for an anisotropic star without cosmological constant our present value $Z \leq 0.65$ seems to be satisfactory. It is to further note that this low value surface redshift is not at unavailable in the literature where Shee et al.~\cite{Shee2016} obtained a numerical value for $Z$ as $0.30$ (also see following Refs.~\cite{Kalam2012,Rahaman2012b,Hossein2012,Kalam2013,Bhar2015} for low valued surface redshift).

\begin{table}
\centering \caption{Numerical values of physical parameters
$nA$, $AR^2$, $A$,$B$, $F$ and $D$ for the different
values of $n$ for $RXJ 1856-37$}\label{Table 1}
{\begin{tabular}{@{}ccccccc@{}}

\hline

$n$ & $nA$ & $AR^2$ & $A(km^{-2})$ & $B$ & $F(km^2)$ & $D$ \\ \hline

$3$ & 0.0128 &  0.1535  & 0.0043 & 0.3623 & 81.0930 & 4.5094 \\ \hline
$4$ &  0.0123 &  0.1109  & 0.0031 & 0.3651 & 81.0968 & 5.8353 \\ \hline
$10$ & 0.0115&  0.04158  & 0.0012 &  0.3700 & 81.0905 & 13.8611 \\ \hline
$100$ & 0.0111 & 0.004009  & 1.1136$\times 10^{-4} $ & 0.3727 & 81.0794 &  134.5882 \\ \hline
$1000$ &  0.0111 & 3.9950$\times 10^{-4}$ & 1.1095$\times 10^{-5} $ & 0.3729 & 81.0877 & 1.3420$\times 10^{3} $\\ \hline
$10000$ &  0.0111 & 3.9920$\times 10^{-5}$&  1.1089$\times 10^{-6}$&  0.3730 & 81.0930 & 1.3418$\times 10^{4}$ \\ \hline
\end{tabular}}
\end{table}

\begin{table}
\centering \caption{Numerical values of physical parameters
$M\left(M_\odot\right)$, $R~(km)$ and $A{R}^{2}$ for the different
values of $n$} \label{Table 3}

{\begin{tabular}{@{}cccccccc@{}}

\hline

& & & $ n=3$ & $n=4$ & $n=10$ & $n=100$ &$n=10000$\\

\hline

Compact stars & $M~(M_\odot)$ & $R~(km)$ & $A{R}^{2}$ & $A{R}^{2}$
& $A{R}^{2}$ & $A{R}^{2}$ & $A{R}^{2}$ \\

\hline RXJ 1856-37 & 0.9042 & 6.002  & 0.1535 & 0.1109 &0.04158 &0.004009&3.9920$\times 10^{-5}$ \\

\hline SAX J1808.4- 3658(SS2) & 1.3238 & 6.35  & 0.3610 & 0.2484 & 0.08643 & 0.00802 & $7.96\times10^{-5}$ \\

\hline SAX J1808.4- 3658(SS1) & 1.4349 & 7.07  & 0.3295 & 0.2283 & 0.0803  & 0.00749 & $7.438\times10^{-5}$ \\

\hline
\end{tabular}}
\end{table}

\begin{table}
	\centering
	\caption{Central density, Surface density and Central pressure for compact star candidate $RXJ 1856-37$ for the above parameter values}\label{Table 2}
	
\begin{tabular}{@{}lrrr@{}} \hline
value & Central Density & Surface density & Central pressure   \\
of $n$ & $gm/cm^{3} $ & $gm/cm^{3}$ & $dyne/cm^{2} $ \\\hline
$3$  & 3.0973$\times 10^{15} $ & 1.4966$\times 10^{15} $ & 3.0722$\times 10^{35} $   \\ \hline
$4$ & 2.8952$\times 10^{15} $ & 1.5455$\times 10^{15} $ & 3.2227$\times 10^{35}$   \\ \hline
$10$ &  2.5791$\times 10^{15} $ & 1.6339$\times 10^{15} $ & 3.4275$\times 10^{35}$   \\ \hline
$100$ & 2.4145$\times 10^{15} $ & 1.6870$\times 10^{15} $ & 3.5212$\times 10^{35}$   \\ \hline
$1000$ & 2.3987$\times 10^{15} $ & 1.6921$\times 10^{15} $ & 3.5294$\times 10^{35}$   \\ \hline
$10000$ & 2.3969$\times 10^{15} $ & 1.6926$\times 10^{15} $ & 3.5283$\times 10^{35}$   \\ \hline
\end{tabular}
\end{table}

(7) In Table 2 and 3 we have calculated the central density, surface density and central pressure as well as mass and radius of different compact stars. It is interesting to note that all the data are fall within the observed range of the corresponding star's physical parameters~\cite{Kalam2012,Rahaman2012b,Hossein2012,Kalam2013,Bhar2015,Maurya2016,Deb2016}. It would be interesting to perform a comparative study with the data of our Table 3 with that of Table 4 of Deb et al.~\cite{Deb2016} where they have prepared the table for the value of $n=3.3$ only for different compact stars whereas in the present study our table includes a wide range of $n=3$ to $n=10000$. Thus, the Table 3 describes behaviour of certain physical parameters for varying $n$. It is observed that the central density decreases with increasing $n$ unlike the surface density which behaves oppositely. However, the central pressure increases with increasing $n$.

The overall observation is that our proposed model satisfies all physical requirements. The entire analysis has been performed in connection to direct comparison of some of the compact star candidates, e.g. $RXJ~1856-37$, $SAX~J~1808.4-3658~(SS1)$ and $SAX~J~1808.4-3658~(SS2)$ which confirms validity of the present model.

\section*{Acknowledgments}
SKM acknowledges support from the authority of University of
Nizwa, Nizwa, Sultanate of Oman. SR is thankful to the Inter-University
Centre for Astronomy and Astrophysics (IUCAA), Pune, India for providing
Visiting Associateship under which a part of this work was carried
out. SR is also thankful to the authority of The Institute of
Mathematical Sciences, Chennai, India for providing all types
of working facility and hospitality under the Associateship scheme.

\end{document}